\pdfoutput=1

\documentclass[11pt]{article}

\newif\ifarxiv
\arxivtrue 

\ifarxiv
    \usepackage{acl}
\else
    \usepackage[review]{acl}
\fi
\usepackage{csquotes}
\usepackage{times}
\usepackage{latexsym}
\usepackage{amsmath}
\usepackage[T1]{fontenc}

\usepackage[utf8]{inputenc}

\usepackage{microtype}

\usepackage{inconsolata}
\usepackage{enumitem}
\usepackage{array}
\usepackage[capitalise]{cleveref}
\usepackage{listings}
\usepackage{graphicx}
\usepackage{longtable}
\usepackage{booktabs}
\usepackage{soul,xcolor}
\usepackage{longtable}
\sethlcolor{yellow}
\usepackage{subcaption}
\usepackage{multirow}
\usepackage{bm}
\usepackage{amsthm}
\usepackage{url}

 \usepackage[most]{tcolorbox}

\newtcolorbox{myquotebox}{
  colback=white!0, 
  colframe=black, 
  rounded corners,
  boxrule=0.5pt, 
  title=Prompt:,
  left=2mm, 
  right=2mm, 
  top=1mm, 
  bottom=1mm 
}
\usepackage{todonotes}
\usepackage{xspace}

\definecolor{lightgrey}{RGB}{158, 158, 158}
\definecolor{goldenrod}{rgb}{0,0,0.8}
\definecolor{deepred}{rgb}{0.6,0,0}
\definecolor{deepgreen}{rgb}{0,0.5,0}
\definecolor{pink}{RGB}{219, 48, 122}
\definecolor{forestgreen}{RGB}{34,139,34}
\definecolor{goldenrod}{RGB}{218,165,32}
\definecolor{sepia}{RGB}{112,66,20}

\crefname{figure}{Fig.}{Figs.}
\crefname{table}{Table}{Tables}
\crefname{appendix}{App.}{App.}
\crefname{section}{§}{§§}
\crefname{equation}{Eq.}{Eqs.}

\newcommand\myparagraph[1]{
\vskip 0.05in 
\noindent{\bf {#1}}}

\title{UsefulBench: Towards Decision-Useful Information as a Target for Information Retrieval}

\author{ 
    Tobias Schimanski\textsuperscript{\rm 1,2},
    Stefanie Lewandowski\textsuperscript{\rm 3},
    Christian Woerle\textsuperscript{\rm 4}, \\
    \textbf{Nicola Reichenau}\textsuperscript{\rm 3},
    \textbf{Yauheni Huryn}\textsuperscript{\rm 4},
    \textbf{Markus Leippold}\textsuperscript{\rm 1,5}  \\
    \textsuperscript{\rm 1}University of Zurich
    \hspace{5mm}
    \textsuperscript{\rm 2}ETH Zurich
    \hspace{5mm} 
    \textsuperscript{\rm 3}score4more GmbH \hspace{5mm} \\
    \textsuperscript{\rm 4}Climate+Tech Think Tank \hspace{5mm}
    \textsuperscript{\rm 5}Swiss Finance Institute (SFI) \\
    \texttt{tobias.schimanski@df.uzh.ch} \\
}

\begin{document}
\maketitle
\begin{abstract}
Conventional information retrieval is concerned with identifying the \textit{relevance} of texts for a given query. Yet, the conventional definition of relevance is dominated by aspects of similarity in texts, leaving unobserved whether the text is truly \textit{useful} for addressing the query. For instance, when answering whether Paris is larger than Berlin, texts about Paris being in France are relevant (lexical/semantic similarity), but not useful. In this paper, we introduce \textit{UsefulBench}, a domain-specific dataset curated by three professional analysts labeling whether a text is connected to a query (\textit{relevance}) or holds practical value in responding to it (\textit{usefulness}). We show that classic similarity-based information retrieval aligns more strongly with \textit{relevance}. While LLM-based systems can counteract this bias, we find that domain-specific problems require a high degree of expertise, which current LLMs do not fully incorporate. We explore approaches to (partially) overcome this challenge. However, \textit{UsefulBench} presents a dataset challenge for targeted information retrieval systems.\footnote{The dataset is available for researchers. For dataset access, please reach out to \textit{stefanie.lewandowski@score4more.eu} and \textit{tobias.schimanski@df.uzh.ch}.}
\end{abstract}

\section{Introduction} \label{sec:introduction}
Modern Large Language Models (LLMs) have become increasingly capable of solving a wide range of tasks like coding and math \citep[e.g.,][]{singh2025openaigpt5card, qwen3, grattafiori2024llama3herdmodels}. Using LLMs for question answering tasks usually comes with concerns about hallucination and outdated information. To mitigate these concerns, a wide range of applications employ retrieval augmented generation (RAG), where an LLM is provided with external context to answer a question \citep{lewis2021retrievalaugmentedgenerationknowledgeintensivenlp}. However, with the emergence of RAG, there also started an ongoing debate about its reliability in various use cases \citep{liu2023evaluatingverifiabilitygenerativesearch, schimanski-etal-2024-towards, xian2025vulnerabilityapplyingretrievalaugmentedgeneration}.

\begin{figure}[t]
    \centering	\includegraphics[width=\columnwidth]{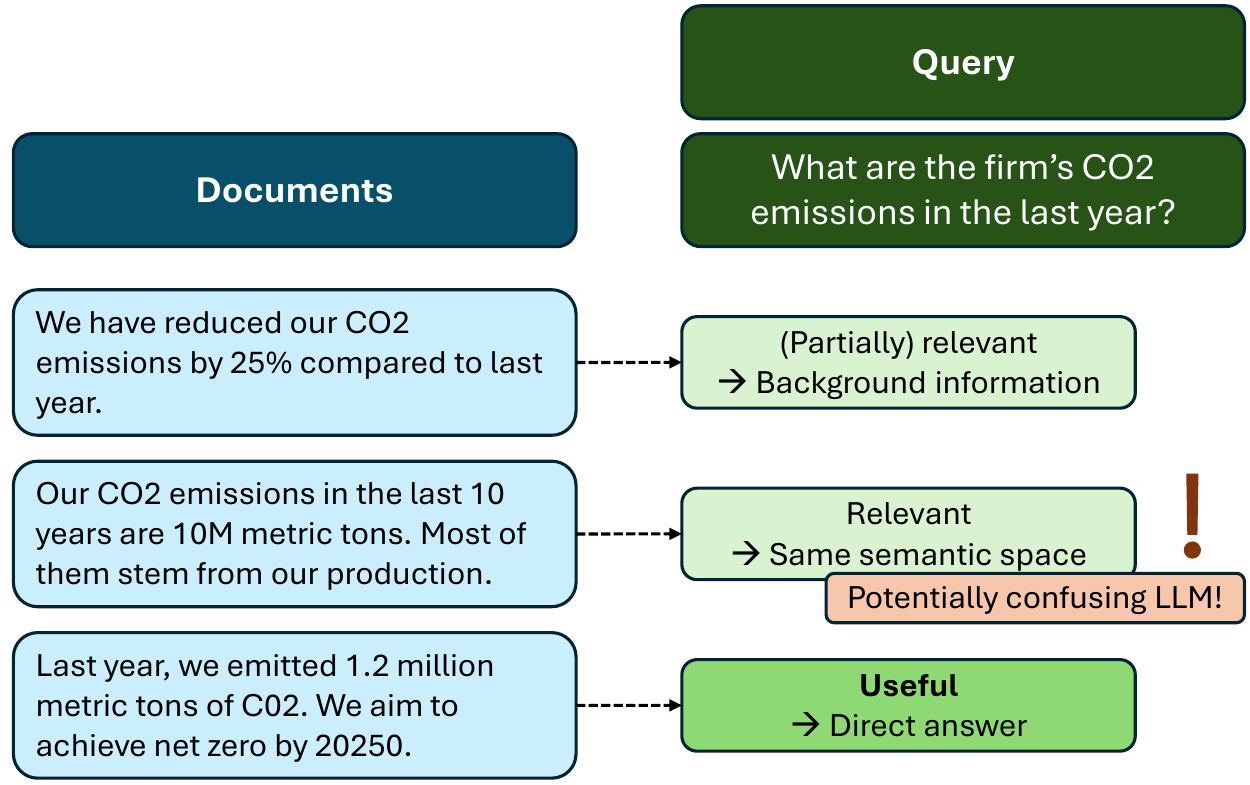}
	\vspace{-1em}
    \caption{Relevance and usefulness examples. Documents can be highly relevant, yet not useful in producing an answer to the query. Highly relevant, yet non-useful documents can harm answer quality \citep[see e.g.,][]{Cuconasu2024, xian2025vulnerabilityapplyingretrievalaugmentedgeneration}.
    }
	\label{fig:example_rel_useful}
    \vspace{-0.5em}
\end{figure}

One major bottleneck in reliable RAG is the retrieval process. Prior research has shown that large contexts can distort QA performance \citep{liu-etal-2024-lost}. At the same time, the highest-ranking documents in the retrieval process, that do not contain an answer or documents with misleading evidence, can majorly distort effective QA performance \citep{Cuconasu2024, xian2025vulnerabilityapplyingretrievalaugmentedgeneration, zeng2026worsezeroshotfactcheckingdataset}. This motivates a differentiation of documents into two categories: \textit{relevant} and \textit{useful}. Relevant documents contain query-related information (may help contextualize an answer), while useful documents contain direct information to respond to the query (see Figure \ref{fig:example_rel_useful}).

Yet, conventional information retrieval systems are leaning towards what we define as relevance. They rank documents according to lexical or semantic similarity \citep[e.g.,][]{Bm25andbeyond_2009, santhanam-etal-2022-colbertv2}, which we confirm in this paper (Section \ref{sec:embeddings_retrieval}). At the same time, prior research has shown that advanced similarity-based embedding systems can fail with simple queries, due to non-avoidable theoretical limitations \citep{weller2025theoreticallimitationsembeddingbasedretrieval}. To overcome some of these concerns, prior research has suggested using LLM-based relevance-annotators \citep{es-etal-2024-ragas, saad-falcon-etal-2024-ares}, also incorporating degrees of relevance \citep{ni-etal-2025-diras}.

However, to the best of the authors' knowledge, no project has aimed to disentangle relevant and useful documents. For this reason, we introduce \textit{UsefulBench}, a domain-specific dataset that contains documents annotated by professional analysts towards degrees of relevance and usefulness. Specifically, we focus on the domain of sustainability reporting, where queries often cover a wide range of relevant and decision-useful content. For instance, the query "What are a firm's CO2 emissions?" may be widely talked about in a sustainability report, through mentioning their importance, how a firm defines them, or background information on their origin. However, only the concrete CO2 emissions number is \textit{useful} information (see Figure \ref{fig:example_rel_useful}).

\begin{figure*}[t]
    \centering
	\includegraphics[width=1\textwidth]{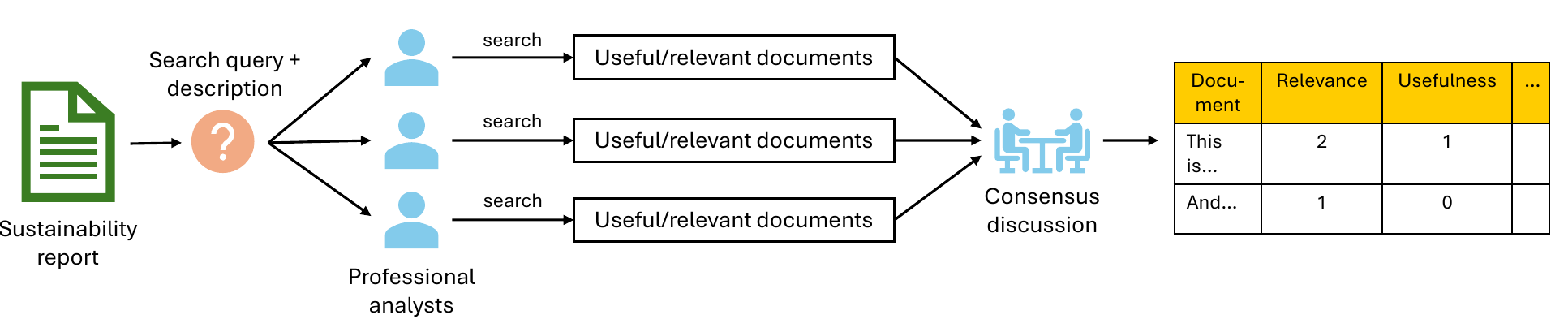}
	\caption{UsefulBench creation pipeline. Three professional analysts search for documents (text passages) that are relevant and useful for a given query and query description. The analysts then consolidate their findings in a consensus discussion to determine the final labels.}
	\label{fig:overview}
 \vspace{-0.5em}
\end{figure*}

To create \textit{UsefulBench}, we employ an expert analysis process by equipping three professional sustainability analysts with real-world queries and query descriptions. The analysts search through firms' sustainability reports and identify relevant and useful documents. Each identified document is annotated to be either non-relevant, partially relevant, or fully relevant, as well as non-useful, partially useful, or fully useful (see Figure \ref{fig:overview}). As a result, we obtain a dataset covering 1,061 annotated documents across 15 sustainability reports using 64 different queries (\textit{UsefulBench-gold}). We also extend \textit{UsefulBench-gold} to a report-level dataset covering all documents in a report with and without relevance and usefulness, and comprise a dataset with 53K report-query-document triplets (\textit{UsefulBench-full}).

Using \textit{UsefulBench}, we show that relevance and usefulness are naturally connected, as expected. However, there remain clear differences. Conventional similarity-based IR systems align more with our definition of relevance. While LLM-based systems can counteract some of this bias and effectively improve in detecting usefulness, we observe an early upper bound of performance. When two professional analysts investigate the misclassifications, they find that LLM-based judgments do not fully incorporate expert knowledge and may misinterpret given query descriptions. We further show that amongst a set of remedy approaches -- from providing examples, refining query descriptions, prompting techniques, and finetuning -- there are often only partial gains, trading off classification against calibration. The most promising approaches directly incorporate expert knowledge in the classification process. However, accurate judgments for these domain-specific queries remain a challenge. 

Collectively, our contributions are:
\begin{itemize}[noitemsep, topsep=0pt]
    \item We introduce \textit{UsefulBench}, a dataset annotated by three professional analysts to differentiate relevance and usefulness of documents to a query.
    \item We show that, while relevance and usefulness are entangled, there are clear differences, and conventional IR systems align more towards relevance.
    \item LLM-based systems integrating expert knowledge can serve as a partial remedy, but ultimately suffer from a lack of expert knowledge.
\end{itemize}

\section{Literature Background}
\textbf{Retrieval and relevance.} Information retrieval (IR) has traditionally focused on identifying documents that are \textit{relevant} to a query. In classical IR, relevance is approximated through lexical matching (e.g., in BM25 \citep{Bm25andbeyond_2009}). More recent neural retrievers replace lexical overlap with semantic similarity, for example, in DPR \citep{karpukhin-etal-2020-dense} or ColBERTv2 \citep{santhanam-etal-2022-colbertv2}. Benchmarks such as BEIR have standardized the comparison of lexical and dense retrieval models across domains \citep{thakur2021beirheterogenousbenchmarkzeroshot}. However, across these settings, relevance is still largely operationalized as topical or semantic relatedness, which does not guarantee that a document contains the information needed to answer a query. This limitation has become more pronounced with recent work showing theoretical and empirical limitations of embedding-based retrieval for certain query types \citep{weller2025theoreticallimitationsembeddingbasedretrieval}.

\noindent\textbf{Retrieval for RAG.} The rise of retrieval-augmented generation (RAG) has made this limitation more consequential, because retrieval errors directly affect answer generation. Prior work shows that passages that are topically related but do not contain the answer can harm downstream performance \citep{Cuconasu2024, zeng-etal-2025-towards}, while long contexts make it harder for language models to identify and use the right evidence \citep{liu-etal-2024-lost}. Similar performance decreases have been shown when the evidence contains contradictory or malicious documents \citep{xian2025vulnerabilityapplyingretrievalaugmentedgeneration, zeng2026worsezeroshotfactcheckingdataset}. At the same time, RAG systems suffer when no useful evidence for QA is present at all \citep{peng-etal-2025-unanswerability}. These challenges are particularly salient in domain-specific settings such as climate and sustainability analysis, where systems must retrieve from long technical reports \citep{Vaghefi2023ChatClimate, ni-etal-2023-chatreport}.

\noindent\textbf{LLM-based relevance annotation.} A related line of work uses LLMs to evaluate or annotate retrieval quality. RAGAS and ARES introduce automated frameworks for evaluating retrieval and generation components in RAG pipelines \citep{es-etal-2024-ragas, saad-falcon-etal-2024-ares}. At the same time, prior work has outlined challenges in domain-specific retrieval \citep{ni-etal-2025-diras, schimanski-etal-2024-climretrieve}. Generally, LLM-based retrieval systems improve retrieval quality at a price of more computational effort \citep{ni-etal-2025-diras, zhang-etal-2025-utility}. LLMs can work out more fine-grained definitions of relevance, which lean towards what we define as usefulness. 

However, no prior work has taken a stance on an active differentiation between documents that are \textit{relevant} to a query and documents that are \textit{useful} for answering it. This distinction matters especially in knowledge-intensive domains, where many passages discuss the queried topic broadly, but only a small subset provides actionable, decision-useful evidence.

\section{UsefulBench}
This project aims to differentiate relevance and usefulness in information retrieval processes. Since, to the best of the authors' knowledge, no dataset explicitly distinguishes between these two concepts, we introduce \textit{UsefulBench}. 

\subsection{Data Creation} \label{sec:data_creation}
At the core of \textit{UsefulBench} is the distinction between the conventional understanding of relevance and the concept of usefulness. Relevance is typically defined through lexical or semantic similarity between a query and a document. Usefulness, in contrast, reflects whether the information contained in a document contributes directly to answering the query or supporting a decision. We argue that the distinction between relevance and usefulness becomes particularly important in domain-specific settings that require deep contextual understanding \citep{Szymanski2025, ni-etal-2025-diras}. For this reason, we focus on the sustainability domain. 

Sustainability and climate disclosures are inherently qualitative and complex. Even widely used indicators such as emissions can be interpreted in multiple ways. When simplifying a query to \enquote{What are a firm's CO$_2$ emissions?}, sustainability reports typically contain a large amount of relevant background information, such as descriptions of emission reduction strategies, methodological explanations, or discussions of climate risks. While such passages are clearly relevant, only a small subset of documents contains the exact emission numbers required to answer the query. Consequently, relevant information can overshadow truly useful information (see also Figure \ref{fig:example_rel_useful}). 

To reflect a realistic expert workflow, the dataset is constructed based on a professional sustainability analysis process. Three professional sustainability analysts examine firms' sustainability reports using real-world queries accompanied by descriptions (Figure \ref{tab:criteria_description}). Examples include information regarding energy efficiency, carbon footprints, or green IT. Overall, the analysts examine 15 sustainability reports of firms from 10 different industries. Each report is analyzed with, on average, 20.5 queries (standard deviation: 8). This follows a structure where each company is evaluated based on seven industry-specific and up to 14 industry-agnostic queries. 
For each query, the analysts read the entire report and identify passages that contain potentially relevant or decision-useful information. On average, 3.3 documents are annotated per query.

The analysts follow the definitions below when retrieving passages (the analysts call a query \enquote{criteria}; detailed prompts in Figure \ref{fig:relevance_prompt} and \ref{fig:usefulness_prompt}): 
\begin{itemize}[noitemsep, topsep=0pt]
    \item \textit{(High) Relevance}: The information has a strong and direct connection to the criteria. It contains specific keywords, concepts, or themes that are central to the analysis of the criteria. The content is explicitly describing the criteria or highly related and on-topic.
    \item \textit{(High) Usefulness}: The information provides significant practical value. It contains clear, actionable insights, specific numbers, concrete achievements, solutions, or detailed plans that can be directly used. The content offers tangible value for understanding the company's efforts related to the criteria.
\end{itemize}

These definitions reflect the central distinction of this paper. While usefulness generally implies relevance, the reverse does not necessarily hold: information can be highly relevant yet provide little value for answering the query. Beyond binary labels, the annotation scheme allows for partial relevance and partial usefulness. This follows prior work that incorporates fine-grained annotations in domain-specific retrieval tasks \citep[e.g.,][]{schimanski-etal-2024-climretrieve, ni-etal-2025-diras}.

Each analyst independently identifies relevant passages (i.e., documents) in the reports. Afterwards, all identified passages are discussed in a consensus meeting. In cases of disagreement, the analysts jointly review the document and agree on a final label. Thus, we also record how many analysts initially retrieved each document. As a side product of the annotation process, the analysts also tag whether the document speaks about a target, an action, a solution, or background information regarding the query (see Table \ref{tab:action_sol_target_background_description} for details).

As a result, the professional analysts curate a dataset containing 1,061 annotated query–document pairs, named \textit{UsefulBench-gold}. Since each report is read, and the documents are annotated by three professional analysts, we can also derive a report-level dataset that contains the gold labels as relevant/useful documents per query and the rest as non-relevant/useful. We dub this dataset with over 53K report-query-document triplets \textit{UsefulBench-full}. 

\subsection{Data Description}

\begin{table}[t]
\centering
\begin{tabular}{c|ccc}
\hline
          & \multicolumn{3}{c}{usefulness} \\
relevance & 0        & 1        & 2        \\ \hline
0         & 167      & 0        & 0        \\
1         & 87       & 133      & 22       \\
2         & 9        & 243      & 400      \\ \hline
\end{tabular}
\caption{Joint distribution of relevance and usefulness labels in \textit{UsefulBench-gold}.}
\label{tab:label_distribution}
\vspace{-0.5em}
\end{table}

Table \ref{tab:label_distribution} presents the joint distribution of relevance and usefulness labels in \textit{UsefulBench-gold}. As expected, there is a substantial overlap between the two concepts. Because the analysts explicitly search for decision-useful information (as in a real-world workflow), the majority of retrieved documents are highly useful and therefore also highly relevant.

Similarly, partially relevant documents are frequently partially useful, while non-relevant documents are predominantly not useful. This reflects the natural hierarchy between the two concepts: usefulness typically requires a certain level of relevance. However, the dataset also contains a meaningful number of examples where relevance and usefulness diverge. In particular, 21.9\% of the dataset consists of documents that are highly relevant to the query (relevance = 2) but only partially useful (usefulness = 1). These passages often provide contextual information, explanations, or qualitative discussion related to the query without containing the specific information required to answer it (see Appendix \ref{sec:edge_cases_annotation} for edge case discussions).

These misalignments between relevance and usefulness are the central focus of this work, raising the question: \textit{can information retrieval systems reliably distinguish between documents that are merely relevant and those that are truly useful for answering a query?}

\begin{table*}[htbp]
\centering
\begin{tabular}{llcccccc}
\toprule
model & ground truth & ece $\downarrow$ & brier $\downarrow$ & auroc $\uparrow$ & f1 $\uparrow$ & f1\_1 $\uparrow$ & f2\_2 $\uparrow$ \\
\midrule

\multirow{2}{*}{gpt-4.1-nano}
& relevance  & 0.298 & 0.208 & 0.900 & 0.587 & 0.932 & 0.656 \\
& usefulness & 0.474 & 0.385 & 0.777 & 0.424 & 0.780 & 0.226 \\
\midrule

\multirow{2}{*}{gpt-4.1-mini}
& relevance  & \underline{0.171} & \underline{0.125} & \underline{0.927} & \textbf{0.652} & \textbf{0.955} & \underline{0.775} \\
& usefulness & \underline{0.216} & \underline{0.180} & \underline{0.836} & \underline{0.640} & \textbf{0.914} & \underline{0.634} \\
\midrule

\multirow{2}{*}{gpt-4.1}
& relevance  & \textbf{0.097} & \textbf{0.088} & \textbf{0.932} & \underline{0.649} & \underline{0.954} & \textbf{0.834} \\
& usefulness & \textbf{0.170} & \textbf{0.160} & \textbf{0.852} & \textbf{0.659} & \underline{0.911} & \textbf{0.704} \\

\bottomrule
\end{tabular}
\caption{Classification and calibration performance of GPT-4.1 models for relevance and usefulness prediction. Best performing models are marked in \textbf{bold} and second-best in \underline{underline}.}
\label{tab:classification_baseline_result}
\vspace{-0.5em}
\end{table*}

\section{Differentiating Relevance and Usefulness}

This section investigates whether current retrieval approaches can distinguish between \textit{relevance} and \textit{usefulness}. We evaluate embedding-based retrieval systems and Large Language Models (LLMs) on \textit{UsefulBench}. Primary experiments are conducted on \textit{UsefulBench-gold}, which allows classification and calibration evaluation. Besides, \textit{UsefulBench-full} represents a realistic retrieval setting.

\subsection{Experimental Setup}

We evaluate model performance in two complementary settings: \textit{classification and calibration} and \textit{information retrieval}.

\paragraph{Classification and calibration.}
In the primary setting, models are tasked with predicting either the \textit{relevance} or \textit{usefulness} label of a given (query, document) pair. Predictions follow the three-point annotation scale (0--2). We evaluate performance using macro F1 across all classes (f1), as well as binary F1 scores that distinguish between (i) non-relevant/useful vs.\ relevant/useful documents (f1\_1) and (ii) highly relevant/useful vs.\ all other documents (f1\_2).

To assess calibration, we extract token-level probabilities for the predicted labels and compute confidence-weighted scores. Specifically, let $\hat{y} \in \{0,1,2\}$ denote the predicted label and $p(\hat{y})$ the corresponding token-level probability. We define the confidence-weighted prediction as
\begin{equation}
\tilde{y} = \hat{y} \cdot p(\hat{y}).
\end{equation}
These scores are evaluated using Expected Calibration Error (ECE), Brier Score, and AUROC, after min-max scaling to the interval $[0,1]$. This setup follows prior work \citep{ni-etal-2025-diras, kadavathLanguageModelsMostly2022,tianJustAskCalibration2023}.

We implement this setup using the \texttt{gpt-4.1} family of models \citep{openai2024gpt4technicalreport}, specifically \texttt{gpt-4.1-nano}, \texttt{gpt-4.1-mini}, and \texttt{gpt-4.1}. These models differ in size and capability, allowing us to study trade-offs between efficiency and predictive performance. Prompts follow the general definitions and are displayed in Figure \ref{fig:relevance_prompt} and \ref{fig:usefulness_prompt}.

\paragraph{Information retrieval.}
In the secondary setting, models are evaluated on their ability to retrieve and rank relevant or useful documents from the full report corpus for a given query using \textit{UsefulBench-full}. Documents are ranked based on predicted scores $\tilde{y}$, and retrieval performance is primarily assessed using nDCG@k (for $k \in \{5,10,20\}$), which captures both ranking quality and position sensitivity. This is important since our labels capture graded degrees of relevance/usefulness. We further report recall@k, precision@k, and F1@k to provide a comprehensive view of retrieval effectiveness.

We evaluate both embedding-based and LLM-based retrieval approaches. Embedding-based methods include BM25 \citep{Bm25andbeyond_2009}, dense retrieval using BGE-M3 \citep{chen-etal-2024-m3}, reciprocal rank fusion of BM25 and BGE-M3, and a hybrid retrieval approach with reranking of the top-$k=100$ candidates using BGE-Reranker-Large \citep{bge_embedding}, as well as OpenAI embeddings text-embeddings-3-small and text-embeddings-3-large. To ensure consistency across settings, we also use the same \texttt{gpt-4.1} family of models for retrieval by scoring and ranking documents directly based on their predicted relevance or usefulness.

\subsection{Classification and Calibration}
We begin by evaluating LLM-based approaches for predicting relevance and usefulness labels. Table \ref{tab:classification_baseline_result} reports results across the \texttt{gpt-4.1} model family. A first key pattern is that calibration improves consistently with model size. Larger models achieve substantially lower calibration errors. This indicates that predicted confidence scores become increasingly informative as model size grows.

In contrast, improvements in classification performance are more nuanced. While larger models achieve the best overall macro F1 scores, differences across models remain relatively moderate -- especially for relevance. A more differentiated picture emerges when considering binary distinctions. Smaller models, such as \texttt{gpt-4.1-mini}, perform slightly better in distinguishing non-relevant/useful (0) from relevant/useful (1--2) documents. In contrast, the largest model, \texttt{gpt-4.1}, performs best in identifying highly relevant/useful documents (2 vs.\ 0--1). 

Increased model size seems to primarily improve performance at the upper end of the label distribution. Larger models are more effective at identifying documents that contain highly decision-useful information, whereas smaller models are competitive at coarse filtering between irrelevant and relevant content. The greatest improvements concentrate on usefulness predictions, where more expert knowledge is necessary.

\begin{figure}
    \centering
    \includegraphics[width=1\linewidth]{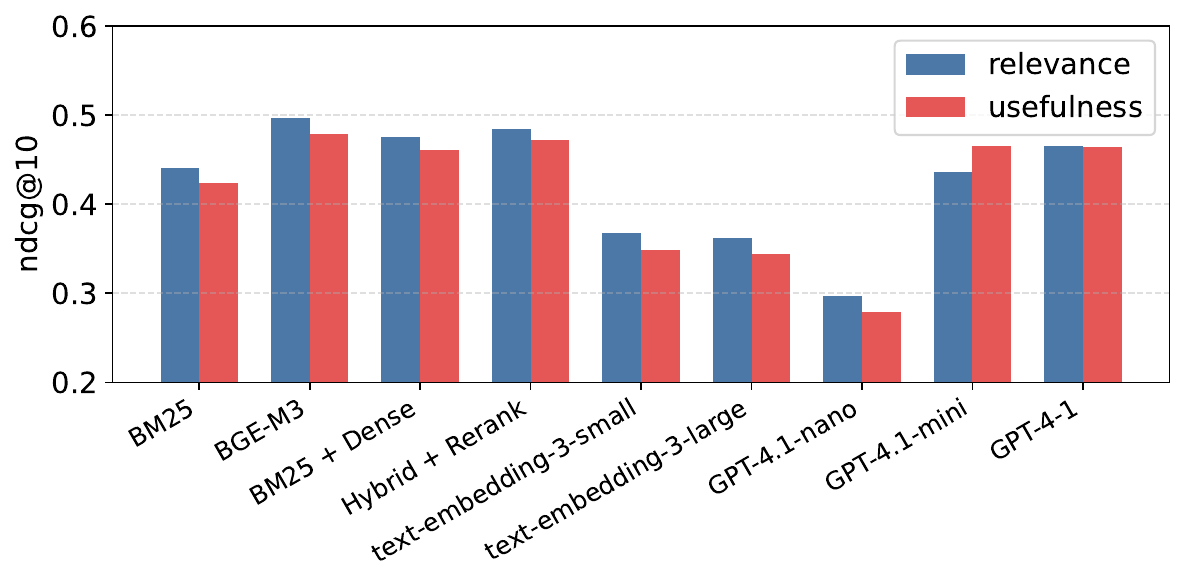}
    \caption{nDCG@10 comparison of embedding and LLM-based rankings.}
    \label{fig:ndcg@20_comparison}
    \vspace{-0.5em}
\end{figure}

\subsection{Information Retrieval}\label{sec:embeddings_retrieval}
Next, we evaluate retrieval performance across embedding-based and LLM-based approaches. Table \ref{tab:ranking_result} and Figure \ref{fig:ndcg@20_comparison} report ranking results across all models.

A first key finding is that classical retrieval systems, including BM25, dense retrieval (BGE-M3), and hybrid approaches, consistently perform better when evaluated on \textit{relevance} than \textit{usefulness} -- consistent with the notion that classical information retrieval systems align with relevance. Turning to LLM-based retrieval, the performance of directly prompting for usefulness is higher than the relevance performance. Hence, LLM-based ranking seems to offer merit in more effectively addressing usefulness. However, gains plateau within the \texttt{gpt-4.1} model family. While moving from smaller nano to mini yields clear improvements, the difference between \texttt{gpt-4.1-mini} and the \texttt{gpt-4.1} model is negligible. At the same time, performance does not improve compared to classical methods. Prompting \texttt{gpt-4.1} is at most (depending on the metric) the second-best IR system after dense retrieval (BGE-M3) (see also Table \ref{tab:ranking_result}).

This pattern suggests the presence of an upper bound for LLM-based retrieval performance driven by the ability to integrate expert-level knowledge. Once a certain level of reasoning and domain understanding is reached, additional model capacity yields limited improvements. This finding is consistent with the classification results, where performance differences between \texttt{gpt-4.1-mini} and \texttt{gpt-4.1} are also relatively small. It also mirrors results of prior work showing that LLMs may lack the in-depth understanding required for complex, knowledge-specific tasks \citep[e.g.,][]{Szymanski2025, ruan2025expertlongbenchbenchmarkinglanguagemodels, chen2025scienceagentbenchrigorousassessmentlanguage, wang-etal-2025-task}.

\subsection{Failure Modes of LLMs}
\begin{table}[t]
\centering
\small
\begin{tabular}{lcccc}
\toprule
 & A & B & C & Disagreement \\
\midrule
Analyst 1 & 30.0\% & 6.0\% & 64.0\% & -- \\
Analyst 2 & 30.0\% & 8.0\% & 62.0\% & -- \\
Agreement   & 28.0\% & 6.0\% & 60.0\% & 6.0\% \\
\bottomrule
\end{tabular}
\caption{Error categories assigned by two analysts for 50 LLM misclassifications: ambiguous description/expert knowledge gap (A), human annotation error (B), and model error (C).}
\label{tab:misclassification_error_analysis}
\vspace{-0.5em}
\end{table}

To ground these results and uncover avenues for improvement, we investigate the nature of misclassifications. We let two of the professional analysts independently assess 50 disagreements between LLM predictions and human labels. For each case, they classify whether the discrepancy arises from (A) ambiguity or lack of expert knowledge in descriptions, (B) annotation errors, or (C) clear model errors (see also \ref{fig:missclassi_analysis}). We specifically ask the experts for category (A) to obtain insights on whether the experts used their own knowledge and interpretation skills beyond the query description to solve the task -- thereby directly checking whether more informed query descriptions may be helpful.

The results in Table \ref{tab:misclassification_error_analysis} reveal a strong consensus between annotators, with agreement in 94\% of cases. Most misclassifications are attributed to clear model errors (analyst 1: 64\%, analyst 2: 62\%), followed by ambiguous cases requiring expert interpretation (both analysts: 30\%). Annotation errors are comparatively rare (6--8\%). The presence of ambiguity in roughly one-third of cases highlights that there is indeed a set of problems where clearer expert descriptions could help the performance. By extension, even clear model errors may arise from wider misinterpretations of the query descriptions. Judgements on relevance and usefulness often integrate expert knowledge that is inaccessible to the model, possibly implicit. For examples of ambiguity, see Appendix \ref{app:examples_missclassi}.

Overall, the analysis suggests that while a large share of errors reflects genuine model limitations, a non-negligible portion arises from a lack of expert know-how in interpreting the descriptions. 

\section{Ablations}
Following these results, we run a set of ablations to check whether we can improve the performance of LLM-based IR systems or solidify our conclusion that expert knowledge is hard to integrate into these models. 
\begin{figure}
    \centering
    \includegraphics[width=1\linewidth]{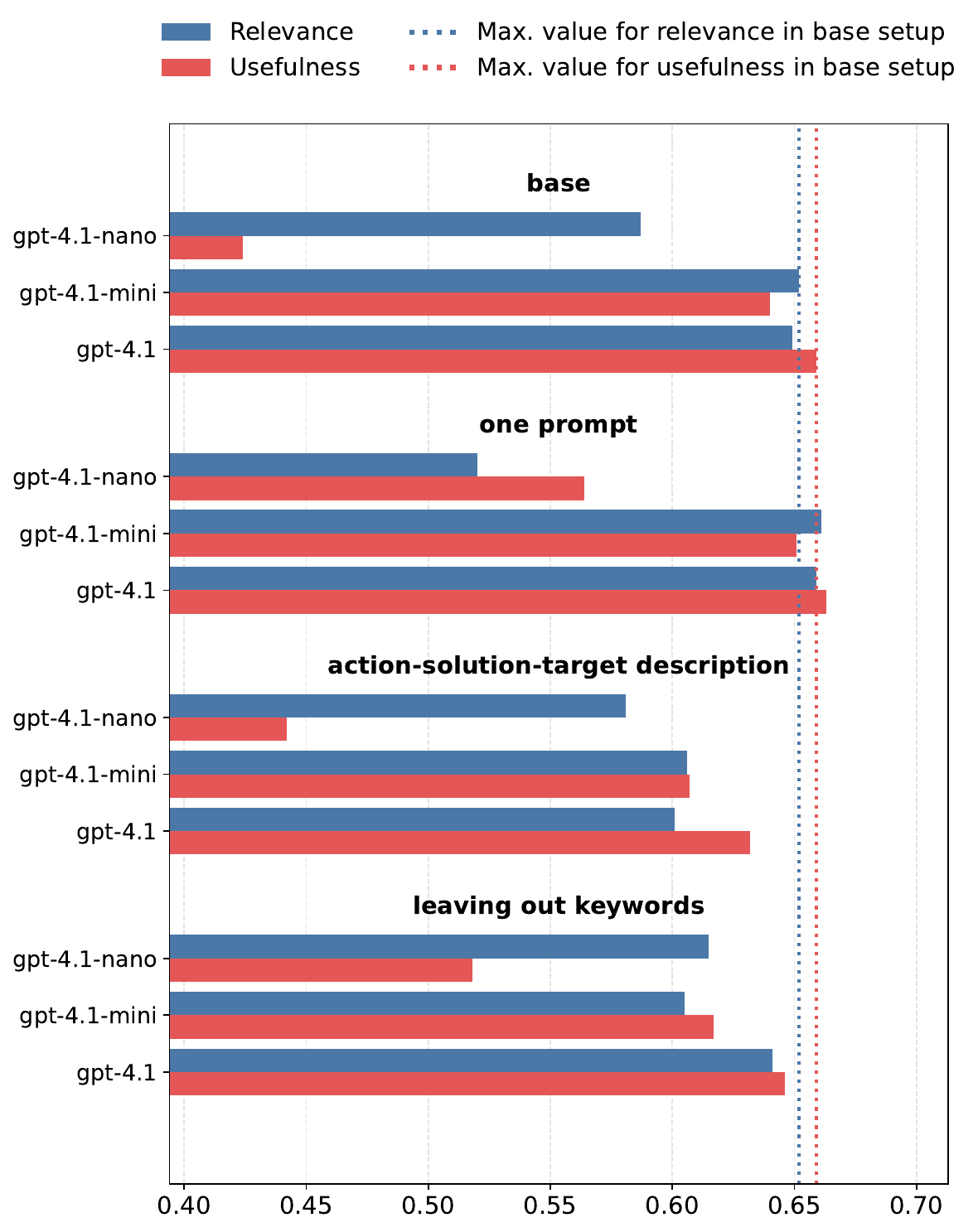}
    \caption{F1 score comparison of the ablations using one prompt for relevance and usefulness, extending descriptions with actions, solution and target definitions, and leaving out keywords. ECE results in Figure \ref{fig:abalations_full_results_ece}.}
    \label{fig:abalations_full_results}
    \vspace{-0.5em}
\end{figure}

\subsection{Multitask Prompting}
In our first ablation, we include both relevance and usefulness definitions in one prompt (Figure \ref{fig:relevance_usefulness_prompt}). Prior research has shown performance improvements when multiple tasks are integrated in one prompt \citep{Xiong2026, son-etal-2024-multi-task}.

Compared to the baseline results (Table \ref{tab:classification_baseline_result}), the highest relevance F1-Score increases from 0.652 to 0.661 (about 1.4\%), and the highest usefulness F1-score rises from 0.659 to 0.663 (about 0.6\%; Table \ref{tab:one_prompt_results}). At the same time, ECE and Brier scores stay close to the baseline. 
Overall, the joint prompt leads to modest performance gains, but the magnitude of improvement remains limited (Figure \ref{fig:abalations_full_results}). From a practical perspective, multitask prompting remains attractive because it allows both judgments to be generated in a single call and is therefore more efficient. However, the results suggest that this efficiency gain does not translate into a substantial improvement in task understanding or predictive quality. 

\subsection{Specialised Descriptions for Targets, Actions, and Solutions}
In a second ablation, we test whether making the query descriptions more specific helps the model distinguish between closely related content types. The texts were annotated towards relevance and usefulness, but also whether they refer to a target, action, or solution (see Section \ref{sec:data_creation}). In this experiment, the descriptions are rewritten so that query descriptions are framed specifically as either \textit{targets}, \textit{actions}, or \textit{solutions} (examples in Table \ref{tab:original_refined_descriptions}). This follows prior work showing that expert-informed descriptions can improve retrieval \citep{schimanski-etal-2024-climretrieve, ni-etal-2025-diras}.

Compared to the baseline results (Table \ref{tab:classification_baseline_result}), the specialized setup does not improve overall classification performance (Table \ref{tab:target_action_solution_prompt_results}). For relevance, the best macro F1-score reaches 0.606, below the baseline best of 0.652. For usefulness, the best macro F1-score is 0.632, also below the baseline best of 0.659. Calibration remains similar across settings. Overall, this additional specialization seems to be unhelpful. The change in definition may have been too small or introduced additional expert know-how, which may have even increased differences in understanding between experts and models.

\subsection{Explanations without Keywords}
Another possible point of confusion may be the fact that the explanation also contain keywords (see e.g., Figure \ref{tab:criteria_description}). If the prompt explicitly contains characteristic terms, the model may rely on lexical associations rather than a deeper understanding of whether a document is actually relevant/useful. To test this, we rely solely on full-text descriptions by removing the keywords.

The results in Table \ref{tab:no_keywords_results} suggest that excluding keywords leads to mixed results. Relative to the baseline in Table \ref{tab:classification_baseline_result}, the best relevance and usefulness macro F1 scores slightly decline. A visible gain appears for gpt-4.1-nano on relevance, where macro F1 rises from 0.587 to 0.615. At the same time, calibration improves noticeably: the lowest relevance ECE falls from 0.097 to 0.062, and the lowest usefulness ECE declines from 0.170 to 0.119. However, these gains do not translate into improved ranking results (Table \ref{tab:ranking_result_no_keywords}). These findings indicate that keyword-heavy definitions may introduce a shallow matching bias for smaller models, but have limited effects on overall results.

\subsection{Few-shot Prompting}
\begin{table}[t]
\centering
\small
\begin{tabular}{llccc}
\toprule
ground truth & setup & ece $\downarrow$ & auroc $\uparrow$ & f1 $\uparrow$ \\
\midrule
usefulness & Base       & \textbf{0.173} & 0.829 & 0.605 \\
usefulness & Similar 10 & 0.217 & \textbf{0.836} & \textbf{0.628} \\
\midrule
relevance  & Base       & \textbf{0.111} & 0.912 & \textbf{0.621} \\
relevance  & Similar 10 & 0.132 & \textbf{0.914} & 0.613 \\
\bottomrule
\end{tabular}
\caption{Classification and calibration performance for GPT-4.1 across the Base and Similar 10 few-shot setups, grouped by ground truth. Better results are in \textbf{bold}. Full results are in Table \ref{tab:classification_fewshot_results}.}
\label{tab:classification_fewshot_gpt41_base_similar10}
\vspace{-0.5em}
\end{table}

Next, we study whether few-shot prompting improves the model’s ability to examine relevance and usefulness. For this, as well as for the following experiment, we split the data into a 60\% training and 40\% test set at the report level (see Table \ref{tab:train_test_split}) and construct few-shot examples from the training split only. We consider four setups: \textit{Random 5}, \textit{Random 10}, \textit{Similar 5}, and \textit{Similar 10}, where examples are either randomly sampled or selected based on the semantic similarity of the document under investigation with documents in the train split using \textit{text-embedding-3-small}.

Table \ref{tab:classification_fewshot_gpt41_base_similar10} shows that few-shot prompting improves classification performance, but introduces a trade-off with calibration. Comparing the baseline to \textit{Similar 10} for \texttt{gpt-4.1}, calibration slightly deteriorates: for relevance, ECE increases from 0.111 to 0.132 (and Brier from 0.100 to 0.116), and for usefulness, ECE rises from 0.173 to 0.217 (Brier: 0.172 to 0.188). At the same time, classification performance improves, with the usefulness macro F1 increasing from 0.605 to 0.628. Once again, gains are stronger in usefulness than in relevance classification. These patterns become more pronounced as the number of examples increases. Moving from 5 to 10 examples (both for random and similarity-based sampling) leads to slightly higher F1 scores, but also to a further deterioration in calibration.

\subsection{Fine-tuning on Train Split}
\begin{figure}
    \centering
    \includegraphics[width=1\linewidth]{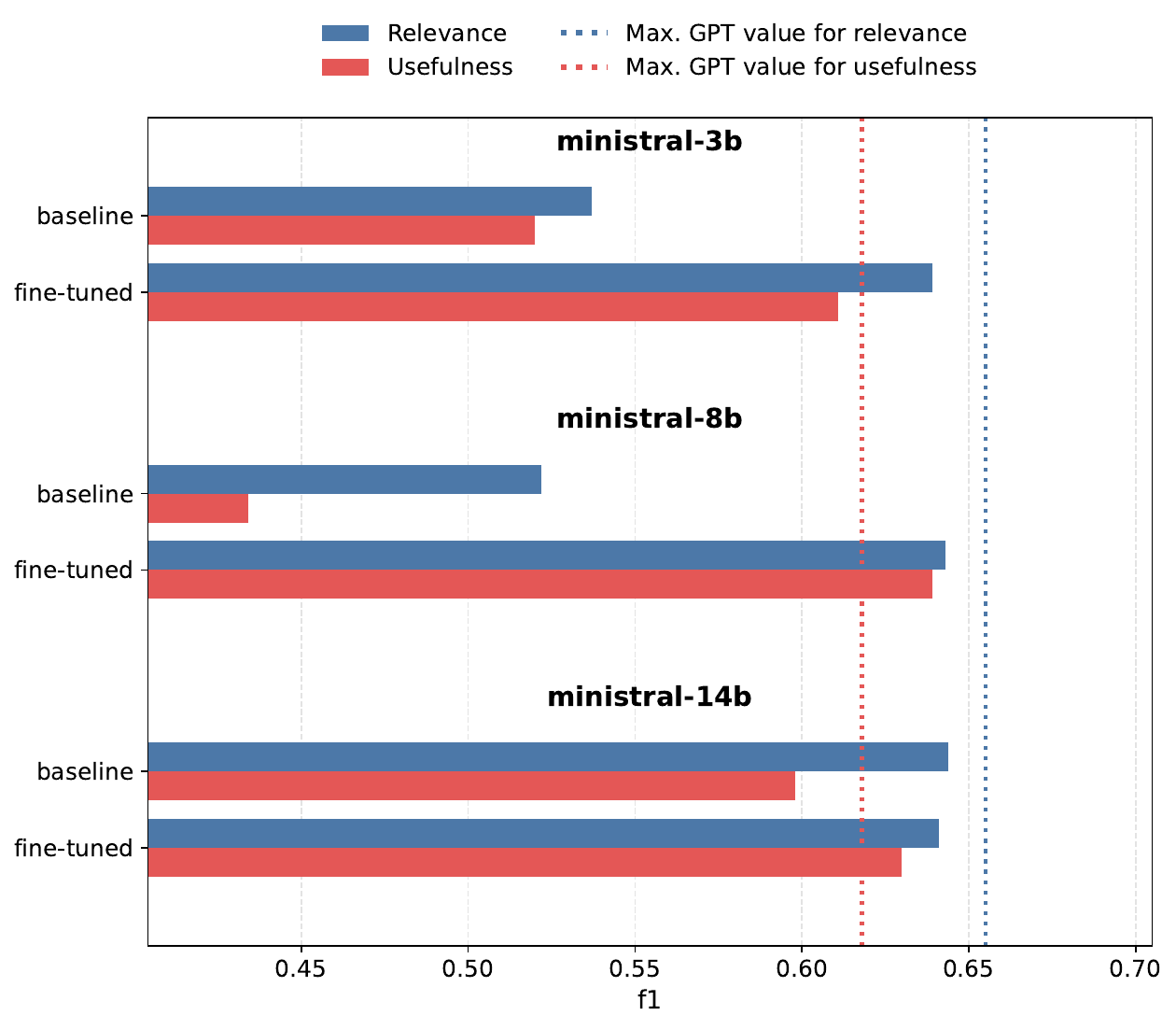}
    \caption{F1 score comparison when fine-tuning Ministral 3b, 8b, and 14b models. ECE results in Figure \ref{fig:finetuing_ece_results}.}
    \label{fig:finetuing_f1_results}
    \vspace{-0.5em}
\end{figure}

Finally, we fine-tune three \texttt{ministral} models of different sizes (3b, 8b, 14b) on the training split to assess whether updating model parameters improves the integration of expert knowledge. Figure \ref{fig:finetuing_f1_results} shows that fine-tuning consistently improves classification performance across all model sizes. The gains are particularly strong for usefulness, where baseline performance is lower. For instance, \texttt{ministral-8b} improves substantially from 0.434 to 0.639 in macro F1 (see Table \ref{tab:classification_mistral_result}). In contrast, relevance shows smaller gains, suggesting that relevance is already relatively well captured in base models. These improvements come, again, with a clear trade-off in calibration. Across all models, ECE and Brier scores increase after fine-tuning. Notably, fine-tuned models reach or exceed GPT-level performance in some cases. For example, \texttt{ministral-8b} achieves a usefulness F1 of 0.639, outperforming \textit{gpt-4.1} (0.605). This suggest that fine-tuning may be the most promising avenue to improve the classification, especially for the target category of usefulness.


\section{Conclusion}
In this paper, we introduce \textit{UsefulBench}, a dataset annotated by professional analysts to differentiate between the relevance and usefulness of documents. Experiments on \textit{UsefulBench} show that relevance and usefulness can be differentiated with LLM-based systems, but it remains challenging to incorporate expert understanding into the process. Improving the access to expert knowledge, e.g., through few-shot prompting and finetuning, proves promising. At the same time, identifying useful documents remains a challenge.

\section{Limitations}
As every project, this project has limitations. First, we concentrate on only one specific domain, namely, sustainability report analysis. While this is a highly knowledge-intensive domain, future work should demonstrate these effects in other domains.

Second, as with every information retrieval dataset, this dataset underlies an annotation selection bias \citep{thakur2021beirheterogenousbenchmarkzeroshot}. This means that annotators may stop labeling at some point, once they deem the information to be sufficient. This may bias the scores to have an earlier upper bound, but preserves the relative order of the scoring approaches.

Third, the core dataset only contains 1,110 datapoints. This originates from the resource intensity of employing three human professional analysts. Future work can extend the dataset quantity to obtain insights on whether these observations remain constant when scaling to more datapoints.

\section*{Ethics Statement}
\myparagraph{Human Annotation}: In this work, all human annotators are professionals in sustainability report analysis who have good knowledge about scientific communication and entailment. They are officially hired and have full knowledge of the context and utility of the collected data. We adhered strictly to ethical guidelines, respecting the dignity, rights, safety, and well-being of all participants. 

\myparagraph{Data Privacy or Bias}: There are no data privacy issues or biases against certain demographics with regard to the data collected from real-world applications and LLM generations.

\myparagraph{Reproducibility Statement}: To ensure full reproducibility, we used all OpenAI models with temperature 0.

\ifarxiv
\section*{Acknowledgements} 
This paper has received funding from the Swiss National Science Foundation (SNSF) under the project `How sustainable is sustainable finance? Impact evaluation and automated greenwashing detection' (Grant Agreement No. 100018\_207800).
\else

\fi

\bibliography{custom}

\appendix

\section{Edge Cases in the Annotation}\label{sec:edge_cases_annotation}
There can be edge cases in the annotation. For instance, a text passage with the labels relevance 2 and usefulness 0 means the text is highly relevant for the query and matches a lot of keywords, e.g. on circular solutions: "We work to create devices that are built with integrity and with reduced waste and carbon impact across their life cycles." (Microsoft 2024). However, the information density, in terms of understanding the company's actions and efforts in this area, is not given. Another example in this domain is marketing language where a company writes about a topic without disclosing any specific information about their engagement/measures. An information-dense text passage (disclosing a specific measure with quantitative data), that does not truly fit the query can have a rating of relevance 1 and usefulness 2. For example, the company writes about a social project supporting women in India with quantitative data, however this is not the core information sought for the query of equal Opportunity, where internal actions targeted at the employees are more relevant. 

\section{Examples in the Misclassification Analysis}\label{app:examples_missclassi}
Companies like to point out certifications they have achieved, with the list of criteria there are five standards that were labeled from the reports. Other criteria were not considered within the labeled profile structure. Due to the separate certification criteria it was not the goal to label e.g. EMAS as part of the energy efficient criteria. Which could be confused by the model because companies often give that as a reason for becoming more energy efficient. Another example where additional knowledge is needed is in business travel, where a companies made internal guideline raising the attractiveness for video calls instead of traveling. Here the video calls had an direct impact on business travel and are an reduction measure. Additional for carbon footprint a company writing about reforestation is a measure with nature based carbon removal. Therefore, highly effective in that category. Clear to the human annotator, but ranked down as less effective by the LLM.

\section{Tables and Figures}

\begin{table*}[htbp]
\centering
\small
\begin{tabular}{p{4cm} p{11cm}}
\toprule
\textbf{Query/Criteria} & \textbf{Description} \\
\midrule

Green IT \& coding &
Green IT \& coding: these are actions and solutions for the transformation to climate-neutral IT, e.g., with climate-neutral data centers, energy-saving hardware and energy-efficient software and algorithms. Typical keywords include: data center waste heat recovery, algorithms, blockchain, energy-efficient IT, energy-saving laptops, green coding, green hosting, green IT, green data center, climate-neutral IT, data center cooling, server cooling, server virtualization, power-saving IT settings, data center heat recovery. \\

Energy efficiency &
Energy efficiency: these are actions and solutions for the transformation to an energy-efficient economy and society. Typical keywords include: motion detector, insulation, efficiency class, energy-efficient cooling, energy saving, energy consumption, heating, ISO 50001, LED, combined heat and power, cooling technology, low-energy building. \\

Renewable energies &
Renewable energies: these are actions and solutions for the transformation to an energy system based on renewable energies. Typical keywords include: biogas, biomass, geothermal energy, offshore, green gas, green electricity, photovoltaics, smart meter, PPA, PV, smart grid, solar, water, wind. \\

Climate-neutral operation &
Climate-neutral operation: these are actions and solutions for the transformation to climate-neutral company operations. Typical keywords include: CO$_2$ compensation, CO$_2$-neutral production, CO$_2$-neutral operation, insulation of company buildings, compressed air optimization, emission reduction, energy efficiency in operations, renewable energy use in operations, climate-neutral production, cooling in operations, conversion to green electricity. \\

Carbon footprint &
Carbon footprint: these refer to the measurement of greenhouse gas emissions in a firm's carbon footprint. Typical keywords include: 2030, 2045, 2050 targets, CO$_2$ equivalent, CO$_2$ emissions, CO$_2$ reduction, CO$_2$e, corporate carbon footprint, direct and indirect emissions, net zero, science-based targets, scope emissions. \\

\bottomrule
\end{tabular}
\caption{Example for climate-related query and their descriptions used for expert annotation and model prompting.}
\label{tab:criteria_description}
\end{table*}

\begin{table*}[htbp]
\centering
\begin{tabular}{p{0.25\textwidth} p{0.73\textwidth}}
\hline
\textbf{Category} & \textbf{Definition} \\
\hline

Action
&
An action is defined as an internal measure the company takes to improve the sustainability of its operations. \\
\hline

Target
&
A target is defined as a goal / not implemented measure the company wants to achieve. Should use forward looking language. \\
\hline

Solution
&
A solution is defined as a product or service the company offers for its end-consumers, that helps them be more sustainable / improve their carbon footprint. \\
\hline

Background Information
&
Background information comprises text passages providing contextual information — such as regulatory frameworks, governance structures, market conditions, or industry trends — that situate but do not describe a company's sustainability activities or frame an action without specifying the action itself. Additionally passages describing concrete company measures that could not be unambiguously assigned to any criterion within the given industry classification, and were retained as background to prevent information loss rather than excluded from the dataset entirely. \\
\hline

\end{tabular}
\caption{Definitions for the classification of actions, solutions, targets and background information.}
\label{tab:action_sol_target_background_description}
\end{table*}

\begin{figure}[ht]
\begin{lstlisting}[frame=single, basicstyle=\ttfamily\scriptsize, xleftmargin=0pt, numbers=none]
"""You are a sustainability analyst, and your task is to rate the relevance of a text towards a given criteria.

<criteria>: '''{description}'''

<text>: '''{text}'''

Please rate the relevance of the text with the given scale:
- 0 = (Not Relevant): The information has little to no connection to the <criteria>. It is off-topic, irrelevant to the specific <criteria>, or consists of general background information that doesn't contribute to the analysis of the <criteria>.
- 1 = (Moderately Relevant): The information has some connection to the <criteria> but is not as direct or central. It may contain related keywords or provide background context, but it doesn't fully address the core concepts of the <criteria>.
- 2 = (Highly Relevant): The information has a strong and direct connection to the <criteria>. It contains specific keywords, concepts, or themes that are central to the analysis of the <criteria>. The content is explicitly describing the <criteria> or highly related and on-topic.

Output only one number in [0, 1, 2]:
"""
\end{lstlisting}
\caption{Prompt for rating the relevance of a document.}
\label{fig:relevance_prompt}
\end{figure}

\begin{figure}[ht]
\begin{lstlisting}[frame=single, basicstyle=\ttfamily\scriptsize, xleftmargin=0pt, numbers=none]
"""You are a sustainability analyst, and your task is to rate the usefulness of a text towards a given criteria.

<criteria>: '''{description}'''

<text>: '''{text}'''

Please rate the usefulness of the text with the given scale:
- 0 = (Not Useful): The information has little to no practical value. It lacks sufficient content, specific actions, solutions or context to be useful for analysis of the <criteria>. It may be a vague statement, or a headline without details.
- 1 = (Moderately Useful): The information offers some practical value but is less specific or actionable. It may provide general context or mention relevant concepts, but it lacks concrete details, specific actions, solutions or data that can be easily utilized to understand the company's sustainability efforts for the <criteria>.
- 2 = (Highly Useful): The information provides significant practical value. It contains clear, actionable insights, specific numbers, concrete achievements, solutions, or detailed plans that can be directly used. The content offers tangible value for understanding the company's efforts related to the <criteria>.

Output only one number in [0, 1, 2]:
"""
\end{lstlisting}
\caption{Prompt for rating the usefulness of a document.}
\label{fig:usefulness_prompt}
\end{figure}

\begin{figure}[ht]
\begin{lstlisting}[frame=single, basicstyle=\ttfamily\scriptsize, xleftmargin=0pt, numbers=none]
Task: Given the criteria, description, and text, as well as the human label and the label by GPT-4.1 (they are in disagreement), the task is to judge why there is a disagreement. Assign one of the 3 labels:
(A) The criteria description or prompt guidelines for usefulness/relevance may be wrong. This means, from a logical point of view, the text is in line with both the human label and the GPT judgment. The definitions may be vague, non-informative, or may not embed the expert knowledge required to correctly judge the text. An adaptation of the definition may solve the conflict. In other words, this is an edge case where disagreement is possible, which would be resolved with expert knowledge.
(B) The human-annotated label assigned to the text is wrong. It is not possible to replicate the judgment of the annotator. The human-annotated label seems to be wrong, and GPT`s annotation is right.
(C) The predicted label by GPT assigned to the text is wrong. It is not possible to understand why the model has gone wrong. The GPT-predicted label is wrong, and the human annotation is right.
\end{lstlisting}
\caption{Instructions for the misclassification analysis.}
\label{fig:missclassi_analysis}
\end{figure}

\begin{figure}[ht]
\begin{lstlisting}[frame=single, basicstyle=\ttfamily\scriptsize, xleftmargin=0pt, numbers=none]
"""You are a sustainability analyst, and your task is to rate the relevance and usefulness of a text towards a given criteria.

<criteria>: '''{description}'''

<text>: '''{text}'''

Please rate the RELEVANCE of the text with the given scale:
- 0 = (Not Relevant): The information has little to no connection to the <criteria>. It is off-topic, irrelevant to the specific <criteria>, or consists of general background information that doesn't contribute to the analysis of the <criteria>.
- 1 = (Moderately Relevant): The information has some connection to the <criteria> but is not as direct or central. It may contain related keywords or provide background context, but it doesn't fully address the core concepts of the <criteria>.
- 2 = (Highly Relevant): The information has a strong and direct connection to the <criteria>. It contains specific keywords, concepts, or themes that are central to the analysis of the <criteria>. The content is explicitly describing the <criteria> or highly related and on-topic.

Please rate the USEFULNESS of the text with the given scale:
- 0 = (Not Useful): The information has little to no practical value. It lacks sufficient content, specific actions, solutions or context to be useful for analysis of the <criteria>. It may be a vague statement, or a headline without details.
- 1 = (Moderately Useful): The information offers some practical value but is less specific or actionable. It may provide general context or mention relevant concepts, but it lacks concrete details, specific actions, solutions or data that can be easily utilized to understand the company's sustainability efforts for the <criteria>.
- 2 = (Highly Useful): The information provides significant practical value. It contains clear, actionable insights, specific numbers, concrete achievements, solutions, or detailed plans that can be directly used. The content offers tangible value for understanding the company's efforts related to the <criteria>.

Output exactly two numbers separated by a comma (relevance,usefulness), e.g., "(1,2)":
"""
\end{lstlisting}
\caption{Prompt for rating the relevance and usefulness jointly.}
\label{fig:relevance_usefulness_prompt}
\end{figure}

\begin{table*}[htbp]
\centering
\tiny
\begin{tabular}{llcccccccccccc}
\toprule
Model & Ground Truth 
& ndcg@5 & ndcg@10 & ndcg@20 
& prec@5 & prec@10 & prec@20 
& rec@5 & rec@10 & rec@20 
& f1@5 & f1@10 & f1@20 \\
\midrule

\multirow{2}{*}{bm25\_score}
& relevance  & 0.414 & 0.440 & 0.474 & 0.462 & 0.533 & 0.628 & 0.157 & 0.107 & 0.068 & 0.201 & 0.154 & 0.111 \\
& usefulness & 0.396 & 0.423 & 0.452 & 0.443 & 0.503 & 0.577 & 0.146 & 0.098 & 0.061 & 0.190 & 0.142 & 0.099 \\
\midrule

\multirow{2}{*}{bge\_m3\_score}
& relevance  & \textbf{0.463} & \textbf{0.497} & \textbf{0.523} & \underline{0.588} & \textbf{0.688} & 0.765 & 0.195 & 0.130 & 0.083 & \textbf{0.255} & 0.192 & 0.134 \\
& usefulness & \textbf{0.450} & \textbf{0.478} & \textbf{0.505} & \underline{0.576} & \textbf{0.642} & 0.710 & 0.188 & 0.120 & 0.076 & \textbf{0.248} & 0.177 & 0.123 \\
\midrule

\multirow{2}{*}{bm25\_dense\_rrf\_score}
& relevance  & 0.455 & 0.475 & 0.519 & 0.523 & 0.579 & 0.734 & 0.179 & 0.118 & 0.079 & 0.230 & 0.169 & 0.129 \\
& usefulness & 0.440 & 0.460 & 0.498 & 0.502 & 0.545 & 0.676 & 0.167 & 0.108 & 0.071 & 0.217 & 0.156 & 0.116 \\
\midrule

\multirow{2}{*}{hybrid\_rerank\_score}
& relevance  & 0.439 & \underline{0.484} & 0.517 & 0.545 & \underline{0.673} & \underline{0.777} & \textbf{0.203} & \textbf{0.144} & \underline{0.091} & \underline{0.254} & \textbf{0.205} & \underline{0.146} \\
& usefulness & 0.427 & \underline{0.472} & \underline{0.502} & 0.529 & \underline{0.643} & \underline{0.731} & \textbf{0.192} & \textbf{0.134} & \textbf{0.084} & \underline{0.244} & \textbf{0.193} & \textbf{0.135} \\
\midrule

\multirow{2}{*}{text-embedding-3-small}
& relevance  & 0.315 & 0.368 & 0.405 & 0.359 & 0.522 & 0.641 & 0.157 & 0.119 & 0.078 & 0.184 & 0.167 & 0.125 \\
& usefulness & 0.291 & 0.349 & 0.383 & 0.333 & 0.491 & 0.588 & 0.144 & 0.109 & 0.072 & 0.170 & 0.154 & 0.114 \\
\midrule

\multirow{2}{*}{text-embedding-3-large}
& relevance  & 0.304 & 0.362 & 0.436 & 0.353 & 0.544 & \textbf{0.807} & 0.171 & 0.128 & 0.088 & 0.197 & 0.180 & 0.143 \\
& usefulness & 0.283 & 0.344 & 0.414 & 0.332 & 0.508 & \textbf{0.747} & 0.157 & 0.119 & 0.081 & 0.182 & 0.167 & 0.132 \\
\midrule

\multirow{2}{*}{gpt-4.1-nano}
& relevance  & 0.228 & 0.297 & 0.360 & 0.324 & 0.536 & 0.734 & 0.137 & 0.116 & 0.087 & 0.171 & 0.171 & 0.140 \\
& usefulness & 0.226 & 0.280 & 0.331 & 0.264 & 0.400 & 0.564 & 0.139 & 0.105 & 0.073 & 0.156 & 0.145 & 0.116 \\
\midrule

\multirow{2}{*}{gpt-4.1-mini}
& relevance  & 0.381 & 0.436 & 0.482 & 0.509 & 0.627 & 0.758 & 0.184 & 0.129 & 0.089 & 0.241 & 0.191 & 0.144 \\
& usefulness & 0.404 & 0.466 & 0.498 & 0.503 & 0.635 & 0.727 & 0.196 & 0.133 & 0.082 & 0.245 & 0.194 & 0.133 \\
\midrule

\multirow{2}{*}{gpt-4.1}
& relevance  & \underline{0.408} & 0.465 & 0.504 & 0.519 & 0.658 & 0.784 & 0.188 & 0.137 & \textbf{0.092} & 0.247 & \underline{0.203} & \textbf{0.150} \\
& usefulness & \underline{0.404} & 0.464 & 0.492 & 0.474 & 0.602 & 0.711 & 0.179 & 0.131 & 0.085 & 0.226 & 0.190 & \underline{0.137} \\

\bottomrule
\end{tabular}
\caption{Retrieval performance of embedding-based and LLM-based methods on UsefulBench-full, evaluated for relevance and usefulness using NDCG, precision, recall, and F1 at $k \in \{5,10,20\}$. Best performing models are marked in \textbf{bold} and second-best in \textit{italic}.}
\label{tab:ranking_result}
\end{table*}

\begin{table*}[htbp]
\centering
\small
\begin{tabular}{llcccccc}
\toprule
Model & Ground Truth & ece & brier & auroc & f1 & f1\_1 & f2\_2 \\
\midrule

\multirow{2}{*}{gpt-4.1-nano}
& relevance  & 0.380 & 0.305 & 0.855 & 0.520 & 0.857 & 0.655 \\
& usefulness & 0.365 & 0.297 & 0.807 & 0.564 & 0.840 & 0.526 \\
\midrule

\multirow{2}{*}{gpt-4.1-mini}
& relevance  & \underline{0.178} & \underline{0.127} & \underline{0.917} & \textbf{0.661} & \textbf{0.955} & \underline{0.774} \\
& usefulness & \underline{0.230} & \underline{0.180} & \textbf{0.853} & \underline{0.651} & \textbf{0.920} & \underline{0.618} \\
\midrule

\multirow{2}{*}{gpt-4.1}
& relevance  & \textbf{0.109} & \textbf{0.094} & \textbf{0.925} & \underline{0.659} & \underline{0.955} & \textbf{0.830} \\
& usefulness & \textbf{0.169} & \textbf{0.154} & \underline{0.842} & \textbf{0.663} & \underline{0.917} & \textbf{0.705} \\

\bottomrule
\end{tabular}
\caption{Classification and calibration performance of GPT models for relevance and usefulness prediction using a single combined prompt that jointly predicts relevance and usefulness. Best performing models are marked in \textbf{bold} and second-best in \underline{underline}.}
\label{tab:one_prompt_results}
\end{table*}

\begin{table*}[htbp]
\centering
\begin{tabular}{p{0.47\textwidth} p{0.47\textwidth}}
\hline
\textbf{Original description} & \textbf{Refined description} \\
\hline

Green IT \& coding: these are actions and solutions for the transformation to climate-neutral it, e.g. with climate-neutral data centers, energy-saving hardware and energy-saving programmed software and algorithms. Typical key words for the criteria are: data center waste heat recovery, algorithms, blockchain, energy-efficient it, energy-saving laptops, green coding, green hosting, green it, green data center, climate-neutral it, data center cooling, server cooling, server virtualization, power-saving it settings, data center heat recovery.
&
Green IT \& coding: these are actions for the transformation to climate-neutral it, e.g. with climate-neutral data centers, energy-saving hardware and energy-saving programmed software and algorithms. \textbf{An action is defined as an internal measure the company takes to improve the sustainability of its operations.} Typical key words for the criteria are: data center waste heat recovery, algorithms, blockchain, energy-efficient it, energy-saving laptops, green coding, green hosting, green it, green data center, climate-neutral it, data center cooling, server cooling, server virtualization, power-saving it settings, data center heat recovery. \\
\hline

Green IT \& coding: these are actions and solutions for the transformation to climate-neutral it, e.g. with climate-neutral data centers, energy-saving hardware and energy-saving programmed software and algorithms. Typical key words for the criteria are: data center waste heat recovery, algorithms, blockchain, energy-efficient it, energy-saving laptops, green coding, green hosting, green it, green data center, climate-neutral it, data center cooling, server cooling, server virtualization, power-saving it settings, data center heat recovery.
&
Green IT \& coding: these are \textbf{targets} for the transformation to climate-neutral it, e.g. with climate-neutral data centers, energy-saving hardware and energy-saving programmed software and algorithms. \textbf{A target is definet as a goal / not implemented measure the company wants to achieve. Should use forward looking language.} Typical key words for the criteria are: data center waste heat recovery, algorithms, blockchain, energy-efficient it, energy-saving laptops, green coding, green hosting, green it, green data center, climate-neutral it, data center cooling, server cooling, server virtualization, power-saving it settings, data center heat recovery. \\
\hline

Carbon footprint: these are measurement of the greenhouse gas emissions in the corporate carbon footprint. Typical key words for the criteria are: 2030, 2045, 2050, co2 equivalent, co2 emissions, co2 reduction, co2eq, co2e, corporate carbon footprint, direct emissions, indirect emissions, carbon footprint, net zero, science-based targets, scope.
&
Carbon footprint: these are \textbf{solutions} \textbf{of} the greenhouse gas emissions in the corporate carbon footprint. \textbf{A solution is defined as a product or service the company offers for its end-consumers, that helps them be more sustainable / improve their carbon footprint.} Typical key words for the criteria are: 2030, 2045, 2050, co2 equivalent, co2 emissions, co2 reduction, co2eq, co2e, corporate carbon footprint, direct emissions, indirect emissions, carbon footprint, net zero, science-based targets, scope. \\
\hline

\end{tabular}
\caption{Examples of original and specialised criterion descriptions used in the target, action, and solution prompt setup. Each refined description reformulates the original criterion for one specific label type and appends a short defining sentence that makes the intended distinction explicit. The table therefore documents the prompt-level transformation from generic criterion descriptions to more fine-grained, type-specific definitions.}
\label{tab:original_refined_descriptions}
\end{table*}

\begin{table*}[htbp]
\centering
\small
\begin{tabular}{llcccccc}
\toprule
Model & Ground Truth & ece & brier & auroc & f1 & f1\_1 & f2\_2 \\
\midrule

\multirow{2}{*}{gpt-4.1-nano}
& relevance  & 0.306 & 0.230 & 0.881 & 0.581 & 0.908 & 0.676 \\
& usefulness & 0.460 & 0.377 & 0.777 & 0.442 & 0.779 & 0.277 \\
\midrule

\multirow{2}{*}{gpt-4.1-mini}
& relevance  & \underline{0.209} & \underline{0.170} & \underline{0.881} & \textbf{0.606} & \textbf{0.924} & \underline{0.744} \\
& usefulness & \underline{0.237} & \underline{0.213} & \underline{0.811} & \underline{0.607} & \textbf{0.884} & \underline{0.617} \\
\midrule

\multirow{2}{*}{gpt-4.1}
& relevance  & \textbf{0.156} & \textbf{0.143} & \textbf{0.894} & \underline{0.601} & \underline{0.923} & \textbf{0.795} \\
& usefulness & \textbf{0.199} & \textbf{0.193} & \textbf{0.829} & \textbf{0.632} & \underline{0.885} & \textbf{0.679} \\

\bottomrule
\end{tabular}
\caption{Classification and calibration performance of GPT models for relevance and usefulness prediction when using specialised criterion definitions for target-, action-, and solution-related content. Best performing models are marked in \textbf{bold} and second-best in \underline{underline}.}
\label{tab:target_action_solution_prompt_results}
\end{table*}

\begin{table*}[htbp]
\centering
\small
\begin{tabular}{llcccccc}
\toprule
Model & Ground Truth & ece & brier & auroc & f1 & f1\_1 & f2\_2 \\
\midrule

\multirow{2}{*}{gpt-4.1-nano}
& relevance  & 0.139 & 0.117 & 0.906 & 0.615 & 0.945 & 0.790 \\
& usefulness & 0.338 & 0.256 & 0.819 & 0.518 & 0.891 & 0.355 \\
\midrule

\multirow{2}{*}{gpt-4.1-mini}
& relevance  & \textbf{0.062} & \textbf{0.082} & \underline{0.922} & 0.605 & \underline{0.944} & \textbf{0.843} \\
& usefulness & \textbf{0.119} & \textbf{0.136} & \textbf{0.856} & \underline{0.617} & \textbf{0.912} & \underline{0.693} \\
\midrule

\multirow{2}{*}{gpt-4.1}
& relevance  & \underline{0.092} & \underline{0.089} & \textbf{0.928} & \textbf{0.641} & \textbf{0.952} & \underline{0.830} \\
& usefulness & \underline{0.152} & \underline{0.157} & \underline{0.848} & \textbf{0.646} & \underline{0.903} & \textbf{0.714} \\

\bottomrule
\end{tabular}
\caption{Classification and calibration performance of GPT models for relevance and usefulness prediction when excluding keywords from the definitions. Best performing models are marked in \textbf{bold} and second-best in \underline{underline}.}
\label{tab:no_keywords_results}
\end{table*}

\begin{figure*}
    \centering
    \includegraphics[width=0.7\linewidth]{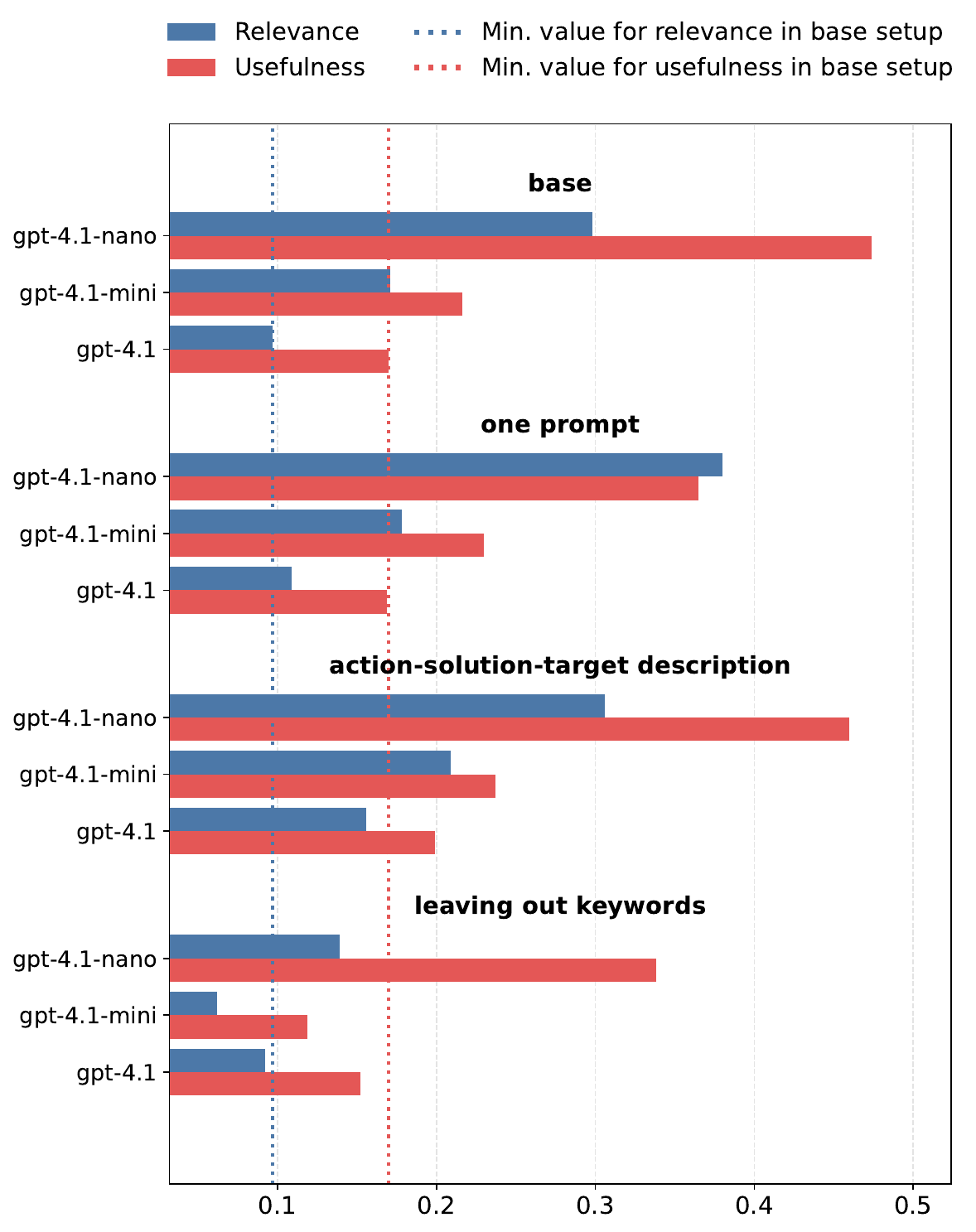}
    \caption{ECE scores comparison of the ablations using one prompt for relevance and usefulness, extending descriptions with actions, solution and target definitions, and leaving out keywords.}
    \label{fig:abalations_full_results_ece}
\end{figure*}

\begin{figure*}
    \centering
    \includegraphics[width=1\linewidth]{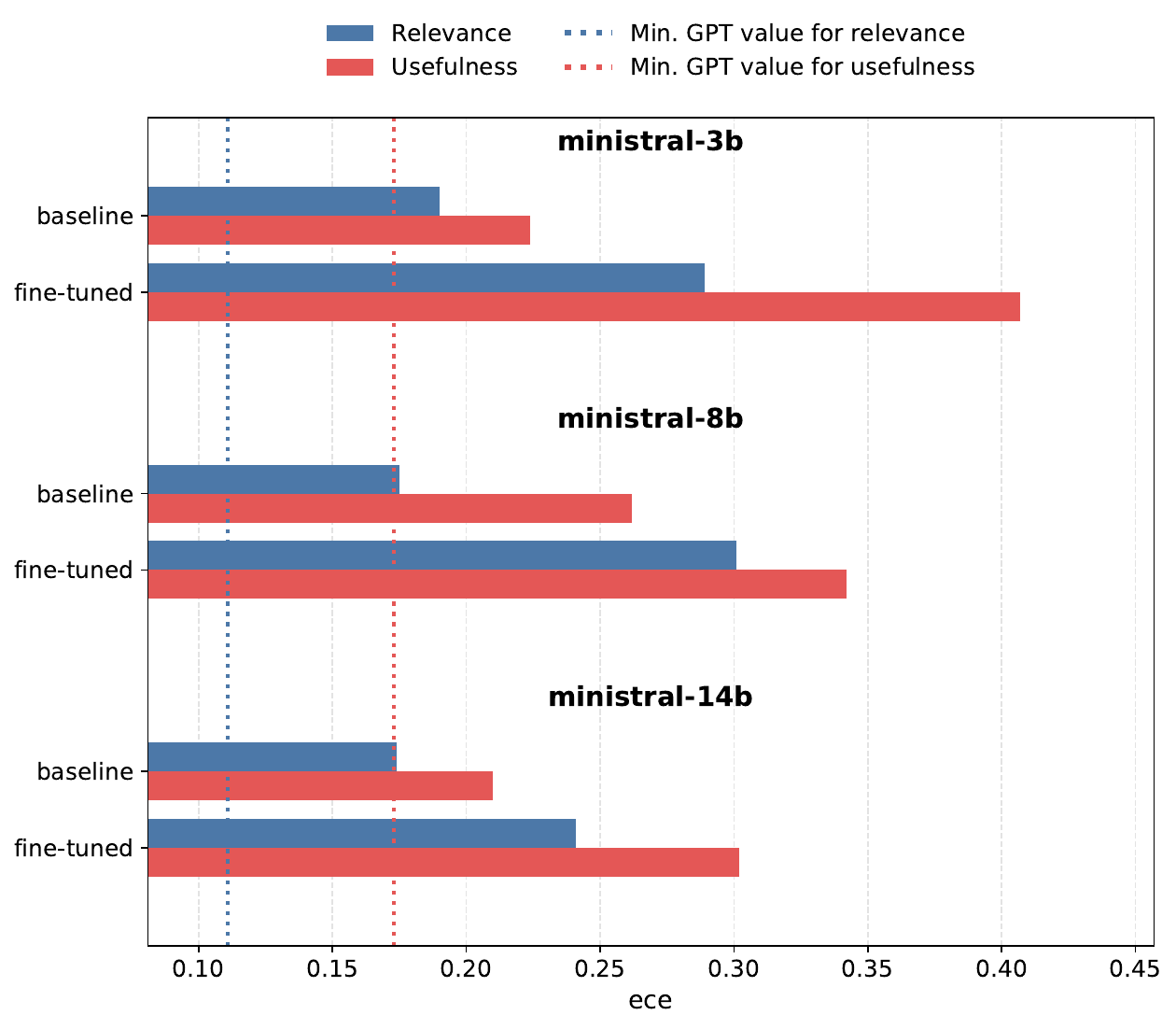}
    \caption{ECE scores comparison when fine-tuning Ministral 3b, 8b, and 14b models.}
    \label{fig:finetuing_ece_results}
\end{figure*}

\begin{table*}[htbp]
\centering
\tiny
\begin{tabular}{llcccccccccccc}
\toprule
Model & Ground Truth 
& ndcg@5 & ndcg@10 & ndcg@20 
& prec@5 & prec@10 & prec@20 
& rec@5 & rec@10 & rec@20 
& f1@5 & f1@10 & f1@20 \\
\midrule

\multirow{2}{*}{bm25\_score}
& relevance  & 0.405 & 0.436 & 0.472 & 0.451 & 0.544 & 0.658 & 0.167 & 0.115 & 0.077 & 0.204 & 0.162 & 0.122 \\
& usefulness & 0.385 & 0.417 & 0.449 & 0.432 & 0.514 & 0.610 & 0.157 & 0.104 & 0.069 & 0.194 & 0.149 & 0.110 \\
\midrule

\multirow{2}{*}{bge\_m3\_score}
& relevance  & 0.442 & 0.477 & 0.508 & \underline{0.569} & \underline{0.671} & 0.762 & 0.198 & 0.134 & 0.089 & \underline{0.253} & 0.195 & 0.142 \\
& usefulness & 0.428 & 0.459 & 0.490 & \textbf{0.558} & 0.636 & 0.716 & 0.191 & 0.124 & 0.082 & 0.246 & 0.182 & 0.131 \\
\midrule

\multirow{2}{*}{bm25\_dense\_rrf\_score}
& relevance  & 0.453 & 0.476 & 0.519 & 0.534 & 0.598 & 0.744 & 0.194 & 0.130 & 0.088 & 0.242 & 0.182 & 0.140 \\
& usefulness & 0.434 & 0.459 & 0.497 & 0.513 & 0.565 & 0.690 & 0.182 & 0.119 & 0.080 & 0.229 & 0.169 & 0.127 \\
\midrule

\multirow{2}{*}{hybrid\_rerank\_score}
& relevance  & \textbf{0.468} & \textbf{0.507} & \textbf{0.539} & 0.561 & \textbf{0.675} & \underline{0.774} & \textbf{0.218} & \textbf{0.151} & 0.096 & \textbf{0.267} & \underline{0.212} & 0.152 \\
& usefulness & \textbf{0.449} & \textbf{0.491} & \textbf{0.521} & 0.544 & \textbf{0.648} & \underline{0.733} & \textbf{0.208} & 0.142 & \underline{0.089} & \textbf{0.257} & 0.201 & 0.142 \\
\midrule

\multirow{2}{*}{text-embedding-3-small}
& relevance  & 0.313 & 0.372 & 0.413 & 0.350 & 0.534 & 0.669 & 0.167 & 0.127 & 0.085 & 0.188 & 0.176 & 0.135 \\
& usefulness & 0.286 & 0.350 & 0.390 & 0.326 & 0.506 & 0.627 & 0.155 & 0.118 & 0.079 & 0.174 & 0.164 & 0.125 \\
\midrule

\multirow{2}{*}{text-embedding-3-large}
& relevance  & 0.310 & 0.365 & 0.438 & 0.350 & 0.533 & 0.796 & 0.183 & 0.135 & 0.094 & 0.202 & 0.184 & 0.149 \\
& usefulness & 0.289 & 0.346 & 0.416 & 0.333 & 0.503 & \textbf{0.747} & 0.172 & 0.127 & 0.087 & 0.191 & 0.173 & 0.139 \\
\midrule

\multirow{2}{*}{gpt-4.1-nano}
& relevance  & 0.234 & 0.301 & 0.371 & 0.343 & 0.542 & 0.775 & 0.156 & 0.130 & 0.099 & 0.184 & 0.182 & 0.156 \\
& usefulness & 0.225 & 0.276 & 0.324 & 0.243 & 0.377 & 0.542 & 0.141 & 0.112 & 0.078 & 0.152 & 0.148 & 0.120 \\
\midrule

\multirow{2}{*}{gpt-4.1-mini}
& relevance  & 0.370 & 0.429 & 0.480 & 0.449 & 0.610 & 0.769 & 0.191 & 0.142 & 0.099 & 0.231 & 0.202 & 0.156 \\
& usefulness & 0.391 & 0.459 & 0.496 & 0.440 & 0.635 & 0.754 & 0.187 & \underline{0.143} & 0.094 & 0.225 & \underline{0.201} & \underline{0.148} \\
\midrule

\multirow{2}{*}{gpt-4.1}
& relevance  & \underline{0.404} & \underline{0.464} & \underline{0.506} & 0.514 & 0.667 & \textbf{0.802} & \underline{0.196} & \underline{0.149} & \textbf{0.104} & 0.251 & \textbf{0.214} & \textbf{0.164} \\
& usefulness & \underline{0.407} & \underline{0.465} & \underline{0.494} & \underline{0.453} & \underline{0.626} & 0.708 & 0.183 & \underline{0.143} & \textbf{0.092} & \underline{0.225} & \textbf{0.202} & \textbf{0.145} \\

\bottomrule
\end{tabular}
\caption{Ranking performance across retrieval and LLM-based scoring models when keyword cues are removed from the criteria explanations. Metrics include NDCG, precision, recall, and F1 at different cutoffs $k \in \{5,10,20\}$, evaluated separately for relevance and usefulness ground truth signals. Best performing models are marked in \textbf{bold} and second-best in \underline{underline}.}
\label{tab:ranking_result_no_keywords}
\end{table*}

\begin{table*}[]
\centering
\begin{tabular}{lll}
\hline
Split & \# reports & \# datapoints \\ \hline
train & 9          & 586           \\
test  & 6          & 475           \\ \hline
\end{tabular}
\caption{Train and test dataset split. The dataset is split along individual reports to prevent leakage of report information from the train to the test split. Approximately 60\% of the dataset is sampled into the train split and 40\% into the test split.}
\label{tab:train_test_split}
\end{table*}

\begin{table*}[htbp]
\centering
\begin{tabular}{lllllllll}
\toprule\midrule
Setup & Model & Ground Truth & ece & brier & auroc & f1 & f1\_1 & f2\_2 \\
\midrule\midrule

\multirow{6}{*}{Base}
& gpt-4.1-nano & relevance  & 0.312 & 0.214 & 0.883 & 0.595 & 0.935 & 0.652 \\
& gpt-4.1-nano & usefulness & 0.479 & 0.392 & 0.767 & 0.383 & 0.773 & 0.179 \\
& gpt-4.1-mini & relevance  & 0.185 & 0.134 & \underline{0.914} & \underline{0.655} & \textbf{0.956} & 0.762 \\
& gpt-4.1-mini & usefulness & 0.215 & 0.191 & 0.808 & \underline{0.618} & 0.900 & \underline{0.642} \\
& gpt-4.1      & relevance  & \textbf{0.111} & \textbf{0.100} & 0.912 & 0.621 & 0.951 & \textbf{0.807} \\
& gpt-4.1      & usefulness & \textbf{0.173} & \textbf{0.172} & 0.829 & 0.605 & 0.900 & 0.653 \\

\midrule\midrule

\multirow{6}{*}{Random 5}
& gpt-4.1-nano & relevance  & 0.263 & 0.217 & 0.851 & 0.550 & 0.899 & 0.756 \\
& gpt-4.1-nano & usefulness & 0.395 & 0.295 & 0.800 & 0.407 & 0.873 & 0.117 \\
& gpt-4.1-mini & relevance  & 0.187 & 0.136 & 0.897 & 0.652 & 0.953 & 0.769 \\
& gpt-4.1-mini & usefulness & 0.215 & 0.186 & 0.802 & 0.619 & \textbf{0.906} & 0.628 \\
& gpt-4.1      & relevance  & \underline{0.119} & \underline{0.101} & 0.913 & 0.629 & 0.955 & 0.799 \\
& gpt-4.1      & usefulness & 0.215 & 0.195 & 0.817 & 0.601 & 0.892 & 0.629 \\

\midrule\midrule

\multirow{6}{*}{Random 10}
& gpt-4.1-nano & relevance  & 0.223 & 0.175 & 0.832 & 0.553 & 0.924 & 0.769 \\
& gpt-4.1-nano & usefulness & 0.329 & 0.279 & 0.757 & 0.536 & 0.854 & 0.561 \\
& gpt-4.1-mini & relevance  & 0.187 & 0.133 & 0.905 & \textbf{0.660} & \underline{0.956} & 0.767 \\
& gpt-4.1-mini & usefulness & 0.218 & 0.186 & 0.808 & 0.617 & \underline{0.905} & 0.631 \\
& gpt-4.1      & relevance  & 0.140 & 0.117 & 0.912 & 0.625 & 0.949 & 0.794 \\
& gpt-4.1      & usefulness & 0.226 & 0.198 & \underline{0.830} & 0.608 & 0.894 & 0.629 \\

\midrule\midrule

\multirow{6}{*}{Similar 5}
& gpt-4.1-nano & relevance  & 0.279 & 0.261 & 0.709 & 0.515 & 0.868 & 0.749 \\
& gpt-4.1-nano & usefulness & 0.356 & 0.320 & 0.718 & 0.531 & 0.817 & 0.601 \\
& gpt-4.1-mini & relevance  & 0.188 & 0.137 & 0.908 & 0.647 & 0.952 & 0.769 \\
& gpt-4.1-mini & usefulness & 0.222 & 0.190 & 0.813 & 0.621 & 0.903 & 0.624 \\
& gpt-4.1      & relevance  & 0.128 & 0.109 & 0.908 & 0.621 & 0.947 & 0.807 \\
& gpt-4.1      & usefulness & \underline{0.214} & 0.198 & 0.822 & 0.616 & 0.886 & \textbf{0.665} \\

\midrule\midrule

\multirow{6}{*}{Similar 10}
& gpt-4.1-nano & relevance  & 0.256 & 0.240 & 0.700 & 0.487 & 0.876 & 0.754 \\
& gpt-4.1-nano & usefulness & 0.313 & 0.302 & 0.658 & 0.462 & 0.821 & 0.583 \\
& gpt-4.1-mini & relevance  & 0.198 & 0.145 & 0.900 & 0.649 & 0.949 & 0.764 \\
& gpt-4.1-mini & usefulness & 0.236 & 0.199 & 0.812 & 0.616 & 0.901 & 0.612 \\
& gpt-4.1      & relevance  & 0.132 & 0.116 & \textbf{0.914} & 0.613 & 0.942 & \underline{0.807} \\
& gpt-4.1      & usefulness & 0.217 & \underline{0.188} & \textbf{0.836} & \textbf{0.628} & 0.895 & 0.663 \\

\bottomrule
\end{tabular}
\caption{Classification and calibration performance across different few-shot setups. Within each ground truth (relevance and usefulness), the best-performing value per column is highlighted in \textbf{bold} and the second-best in \underline{underline}.}
\label{tab:classification_fewshot_results}
\end{table*}

\begin{table*}[htbp]
\centering
\begin{tabular}{llccccccc}
\toprule
model & variant & ground truth & ece $\downarrow$ & brier $\downarrow$ & auroc $\uparrow$ & f1 $\uparrow$ & f1\_1 $\uparrow$ & f2\_2 $\uparrow$ \\
\midrule

\multirow{4}{*}{ministral-3b}
& \multirow{2}{*}{baseline}
  & relevance  & 0.190 & 0.141 & 0.856 & 0.537 & 0.924 & 0.820 \\
& & usefulness & 0.224 & 0.185 & 0.838 & 0.520 & 0.889 & 0.642 \\
\cmidrule{2-9}
& \multirow{2}{*}{fine-tuned}
  & relevance  & 0.289 & 0.171 & 0.897 & 0.639 & 0.946 & \textbf{0.839} \\
& & usefulness & 0.407 & 0.297 & 0.837 & 0.611 & 0.910 & 0.603 \\
\midrule
\multirow{4}{*}{ministral-8b}
& \multirow{2}{*}{baseline}
  & relevance  & \underline{0.175} & \textbf{0.121} & 0.830 & 0.522 & 0.935 & 0.807 \\
& & usefulness & 0.262 & 0.231 & 0.605 & 0.434 & 0.877 & 0.548 \\
\cmidrule{2-9}
& \multirow{2}{*}{fine-tuned}
  & relevance  & 0.301 & 0.189 & \underline{0.910} & \underline{0.643} & \underline{0.948} & \underline{0.828} \\
& & usefulness & 0.342 & 0.240 & \textbf{0.871} & \textbf{0.639} & \underline{0.914} & \underline{0.676} \\
\midrule
\multirow{4}{*}{ministral-14b}
& \multirow{2}{*}{baseline}
  & relevance  & \textbf{0.174} & \underline{0.132} & 0.893 & \textbf{0.644} & \textbf{0.952} & 0.813 \\
& & usefulness & \textbf{0.210} & \textbf{0.182} & 0.824 & 0.598 & 0.907 & \underline{0.680} \\
\cmidrule{2-9}
& \multirow{2}{*}{fine-tuned}
  & relevance  & 0.241 & 0.136 & \textbf{0.928} & 0.641 & \textbf{0.961} & \underline{0.837} \\
& & usefulness & \underline{0.302} & \underline{0.200} & \underline{0.869} & \underline{0.630} & \textbf{0.925} & \textbf{0.686} \\

\midrule
\midrule

\multirow{4}{*}{\textit{gpt-4.1}}
& \multirow{2}{*}{\textit{gpt-4.1-mini}}
  & \textit{relevance}  & \textit{0.185} & \textit{0.134} & \textit{0.914} & \textit{0.655} & \textit{0.956} & \textit{0.762} \\
& & \textit{usefulness} & \textit{0.215} & \textit{0.191} & \textit{0.808} & \textit{0.618} & \textit{0.900} & \textit{0.642} \\
\cmidrule{2-9}
& \multirow{2}{*}{\textit{gpt-4.1}}
  & \textit{relevance}  & \textit{0.111} & \textit{0.100} & \textit{0.912} & \textit{0.621} & \textit{0.951} & \textit{0.807} \\
& & \textit{usefulness} & \textit{0.173} & \textit{0.172} & \textit{0.829} & \textit{0.605} & \textit{0.900} & \textit{0.653} \\

\midrule
\bottomrule
\end{tabular}
\caption{Classification and calibration performance of Ministral models (3b, 8b, 14b) for relevance and usefulness prediction, comparing baseline and fine-tuned variants. Best performing results are marked in \textbf{bold} and second-best in \underline{underline}. GPT results are shown in italic for reference.}
\label{tab:classification_mistral_result}
\end{table*}

\end{document}